\documentclass[aps,twocolumn,showpacs]{revtex4}
\usepackage{epsfig}

\begin{document}

\title{Tumbleweeds and airborne gravitational noise sources for LIGO}
\author{Teviet Creighton}
\affiliation{Department of Physics, University of
Wisconsin---Milwaukee, P.O.\ Box 413, Milwaukee, Wisconsin 53201}
\date{\today}

\begin{abstract}
The test masses in gravitational-wave detectors will be sensitive not
only to astrophysical gravitational waves, but also to the fluctuating
Newtonian gravitational forces of moving masses in the ground and air
around the detector.  These effects are often referred to as gravity
gradient noise.  This paper considers the effects of gravity gradients
from density perturbations in the atmosphere, and from massive
airborne objects near the detector.  These have been discussed
previously by Saulson, who considered the effects of background
acoustic pressure waves and of massive objects moving smoothly past
the interferometer; the gravity gradients he predicted would be too
small to be of serious concern even for advanced interferometric
gravitational-wave detectors.  In this paper I revisit these
phenomena, considering transient atmospheric shocks, and estimating
the effects of sound waves or objects colliding with the ground or
buildings around the test masses.  I also consider another source of
atmospheric density fluctuations: temperature perturbations that are
advected past the detector by the wind.  I find that background
acoustic noise and temperature fluctuations still produce gravity
gradient noise that is below the noise floor even of advanced
interferometric detectors, although temperature perturbations carried
along non-laminar streamlines could produce noise that is within an
order of magnitude of the projected noise floor at 10~Hz.  A
definitive study of this effect may require better models of the wind
flow past a given instrument.  I also find that transient shockwaves
in the atmosphere could potentially produce large spurious signals,
with signal-to-noise ratios in the hundreds in an advanced
interferometric detector.  These signals could be vetoed by means of
acoustic sensors outside of the buildings.  Massive wind-borne objects
such as tumbleweeds could also produce gravity gradient signals with
signal-to-noise ratios in the hundreds if they collide with the
interferometer buildings, so it may be necessary to build fences
preventing such objects from approaching within about 30m of the test
masses.
\end{abstract}
\maketitle

\section{Introduction}
\label{s:gradient-introduction}

Interferometric detectors such as LIGO and VIRGO rely on exquisite
sensitivity to the positions of hanging test masses in order to detect
the perturbations of passing gravitational radiation.  The sensitivity
is so great that the measurements can also be affected by fluctuations
in the local Newtonian gravitational field, which create tiny
accelerations of the mass.  This noise source, known as gravity
gradient noise or Newtonian gravitational noise, is caused by the
near-field gravity of masses moving near the interferometer, and is
not to be confused with the far-field propagating gravitational waves
that the instruments are intended to measure.

Gravity gradient noise has the potential to be quite insidious, since
it cannot be shielded by improvements to the test-mass vibrational
isolation.  The only effective way to eliminate gravity gradient noise
is to eliminate the moving masses that create the perturbing fields.
Fortunately, though, the strongest perturbations to the local
gravitational field are at frequencies well below the detectors'
pass-bands.  Since the proposed terrestrial interferometric detectors
all have sensitivity cutoffs around 3~Hz or higher, we need only worry
about noise sources that can perturb the local gravity field on
timescales less than about 0.3~seconds.  Most of the noise sources
that I consider are motivated by the expected sensitivity of advanced
LIGO interferometers, which were originally projected to have a
low-frequency cutoff around 10~Hz, and instrumental noise of
$S_h\sim2\times10^{-45}\mathrm{Hz}^{-1}$ in the gravitational-wave
signal output at that
frequency~\cite{Abramovici_A:1992}\footnote{Specifically, I will be
using the noise curve in Fig.~10 of~\cite{Abramovici_A:1992}, not
Fig.~7, whose suspension thermal noise is a factor of 3 too small.}.
Although changes in instrumentation technology will modify the
ultimate sensitivity goals of LIGO, this ``standard'' advanced LIGO
noise level is a good reference point when considering new noise
sources.  Also, as pointed out below, gravity gradient noise will make
it difficult to push the detector noise much below this level at
10~Hz, regardless of improvements to the interferometers.

Saulson~\cite{Saulson_P:1984} was the first to estimate the effect of
gravity gradient noise on terrestrial interferometric detectors,
considering the effects of seismic waves passing through the earth and
of sound waves in the air.  In both cases he found the spectral
density of noise in the interferometer path-length difference to be
less than $10^{-39}\mathrm{m}^2\mathrm{Hz}^{-1}$ around 10~Hz,
corresponding to noise in the gravitational-wave signal at levels less
than $10^{-46}\mathrm{Hz}^{-1}$ for a 4~km interferometer.  By
comparison, this is significantly less than the noise floor of
$\sim2\times10^{-45}\mathrm{Hz}^{-1}$ that advanced LIGO
interferometers expect to achieve at 10~Hz.  More recently, a detailed
analysis~\cite{Hughes_S:1998} has been made of seismic gravity
gradient noise; this study indicated that seismic gravity gradient
noise would be within a factor of 2 of the advanced LIGO noise floor
at 10~Hz for most times, and could actually \emph{exceed} this noise
floor during seismically noisy times, making seismic gravity gradients
a significant barrier to improvements in low-frequency sensitivity.
It therefore seems prudent to revisit the issue of \emph{atmospheric}
gravity gradients as well.

In Sec.~\ref{s:atmospheric-pressure-waves} I consider gravity
gradients caused by atmospheric pressure perturbations---the same
noise source considered by Saulson.  Attention is paid, however, to
the effects of the ground, and of buildings that reduce the pressure
noise in the immediate vicinity of the interferometer test masses.
Nonetheless, I find that these tend only to weaken the gravity
gradient noise in the pass-bands of interferometric detectors,
reinforcing the conclusion that this noise source is not of any great
concern.

A much stronger source of high-frequency density perturbations in the
atmosphere is the presence of temperature fluctuations, which are
advected past a detector by the wind.  In
Sec.~\ref{s:atmospheric-temperature-perturbations} I analyze this as a
potential source of gravity gradient noise.  However, I find that
while small-scale temperature perturbations can produce high-frequency
temperature fluctuations at any given point, they do not produce the
same high-frequency fluctuations in the test mass position, since a
given pocket of warm or cool air will affect the test mass
gravitationally over the entire time that it is in the vicinity of the
test mass, which is on the order of seconds.  This produces a cutoff
in the noise spectrum above a few tenths of a Hz.  The presence of
turbulent vortices around the interferometer buildings can increase
the high-frequency component, but still probably not enough to show up
in the gravitational-wave noise spectrum.

In Secs.~\ref{s:shockwaves} and~\ref{s:high-speed-objects} I turn away
from sources of background noise to consider possible sources of
transient gravity gradient signals that might be detected as spurious
events in the gravitational-wave instruments.  Sec.~\ref{s:shockwaves}
extends the analysis in Sec.~\ref{s:atmospheric-pressure-waves} to
look at the effects of atmospheric shockwaves, such as might be
generated by an explosion or supersonic aircraft.  I find that sources
such as these can indeed produce detectable signals that might be
interpreted spuriously as gravitational-wave events.  However, such
signals would easily be vetoed using acoustic monitors outside the
interferometer buildings.

Sec.~\ref{s:high-speed-objects} analyzes the gravity gradients
produced by individual objects, such as wind-borne debris, moving
around outside the interferometer buildings.  Saulson considered this
effect for the case of objects moving with fairly uniform velocity,
but typically, in order to produce significant signal above 3~Hz, an
object's motion must be changing on timescales of less than
0.3~seconds.  In particular, I find that objects \emph{colliding} with
the interferometer buildings produce much stronger signals than
objects simply passing by the buildings.  As an example, tumbleweeds
at the Hanford LIGO facility will be a steady source of spurious
signals in advanced detectors if they are allowed to collide with the
end stations.  Preventing such signals requires shielding a region of
a few tens of meters around the end station, screening any wind-borne
debris that masses more than a few hundred grammes.

Sec.~\ref{s:gradient-conclusions} presents some concluding remarks,
including recommendations to the gravitational-wave experimental
groups and possible directions for further research.

\section{Atmospheric pressure waves}
\label{s:atmospheric-pressure-waves}

Pressure perturbations are the only type of atmospheric gravity
gradient noise considered by Saulson~\cite{Saulson_P:1984}.  The
derivation in this section gives largely the same result as his.

Consider a plane pressure wave with frequency $f$ propagating through
a homogeneous airspace at some sound speed $c$.  If the fractional
pressure change $\delta p/p$ is small, it will induce an adiabatic
density change $\delta\rho/\rho=\delta p/\gamma p$, where
$\gamma\approx1.4$ is the ratio of heat capacities at constant
pressure and constant temperature for air at normal temperatures.  The
gravitational acceleration produced by this wave in the direction of
propagation $\mathbf{e}_z$ is:
\begin{equation}
\label{eq:pressure-gradient-homogeneous}
g_z(t) 	= \int\frac{Gz\delta\rho}{r^3}\,dV
	= \frac{2 G\rho c}{\gamma pf} \delta p(t+1/4f) \; ,
\end{equation}
where $\rho\approx1.3\mathrm{kg}\,\mathrm{m}^{-3}$ and
$p\approx10^5\mathrm{N}\,\mathrm{m}^{-2}$ are the ambient air density
and pressure, and $\delta p(t)$ is the pressure perturbation measured
at the same point as the acceleration is being measured.  By symmetry,
there is no acceleration transverse to the wave.

Now consider sound waves in the vicinity of the interferometer.
First, since the interferometer is only sensitive to motions of the
test mass parallel to the arms, the gravitational acceleration is
reduced by a factor $\cos\theta$, where $\theta$ is the angle between
the propagation direction and the interferometer arm.

\begin{figure}[t]
\epsfig{file=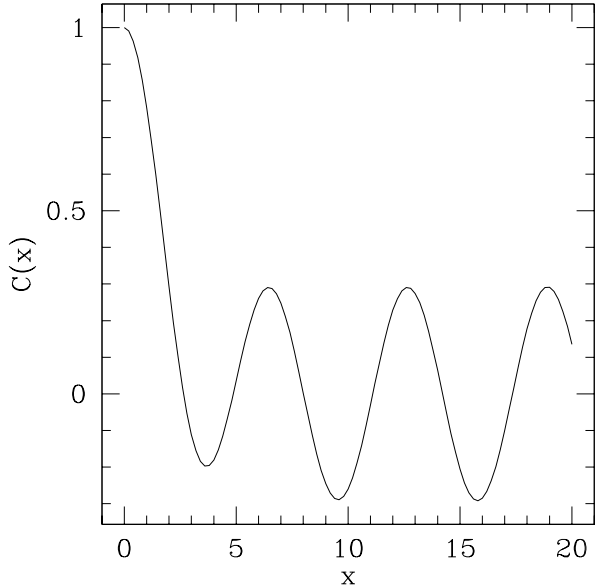,width=3.25in}
\caption{\label{fig:cutoff} Plot of the factor $C(x)$ indicating the
reduction in sonic gravity gradient noise due to setting the density
perturbations to zero within a cylinder of diameter and length
$2r_{\mathrm{min}}$ centered on the test mass (and aligned with the
wave), versus $x=2\pi r_{\mathrm{min}}/\lambda$ where $\lambda$ is the
wavelength.  The oscillatory behaviour above $x=1$ results from the
ends of the cylinder coming in and out of phase with each other.  A
more accurate model of the scattering of sound waves off an
interferometer end station building would modify the behaviour of
$C(x)$ for $x>1$, but probably not change its magnitude
significantly.}
\end{figure}

Second, the interferometer test mass is inside a building, which can
in principle be used to suppress noise within a distance
$r_{\mathrm{min}}$ of the test mass.  Roughly speaking, this results
in a high-frequency cutoff factor $C(2\pi fr_{\mathrm{min}}/c)$, where
the function $C(x)$ depends on the precise shape of the building, the
manner in which it reflects sound waves, and many other factors, but
is normally close to 1 for $x\alt1$.  For instance, if one simply
removes from the volume integral in
Eq.~(\ref{eq:pressure-gradient-homogeneous}) a cylinder with length
and diameter both $2r_{\mathrm{min}}$ aligned with the $z$-axis, then
$C(x)\sim1$ for $x\alt1$, but oscillates with an amplitude of
$\sim0.3$ for $x\agt1$.  This function is shown in
Fig.~\ref{fig:cutoff}.
The constant-amplitude oscillations of $C(x)$ for large $x$ reflect
the assumption that the sound wave has a coherence length much longer
than the building size, so the field between the two ends of the
excluded cylinder is generated almost entirely by the first
half-wavelength beyond each cap.  Realistically, the actual
high-frequency behaviour of $C(x)$ will depend on how the sound waves
bend and scatter around the building; however, this should not change
the order of magnitude of $C(x)$.  The behaviour shown in
Fig.~\ref{fig:cutoff} is therefore probably a good estimate of the
true cutoff function for $x\alt1$, and a reasonable order of magnitude
estimate for $x\agt1$.  For the LIGO end stations, $r_{\mathrm{min}}$
is of order 5~metres, giving $x\sim f/(10\mathrm{Hz})$; the factor
will not be too far off for the frequencies of greatest interest.
More precise estimates would depend on the specific architectural
details of a particular facility.

Third, the interferometer is on the ground, not in homogeneous empty
space.  For simplicity I assume that the waves are almost entirely
reflected off the ground; I will justify this assumption in
Sec.~\ref{ss:ground-absorption}.  In this case the gravity gradient in
directions parallel to the ground contributed by the reflected
wavefront is the same as if the wavefront were extended below ground,
while the pressure perturbations measured by detectors near the ground
(much less than a wavelength) will be doubled.  The acceleration
experienced by an interferometer test mass is therefore:
\begin{equation}
\label{eq:pressure-gradient-interferometer}
g_z(t) = \frac{G\rho c}{\gamma pf} \cos(\theta)
	C(2\pi fr_{\mathrm{min}}/c) \delta p(t+1/4f) \; .
\end{equation}

The gravitational wave signal $h(t)$ in the interferometer is related
to the acceleration of one of the test masses by $\ddot{h}(t)=g(t)/L$,
or in frequency space $\tilde{h}(f)=(2\pi f)^{-2}\tilde{g}(f)/L$,
where $L$ is the length of the interferometer arm.  Thus:
\begin{equation}
\label{eq:pressure-h-tilde}
\tilde{h}(f) = \frac{G\rho c}{4\pi^2\gamma pLf^3} \cos(\theta)
	C(2\pi fr_{\mathrm{min}}/c) i\widetilde{\delta p}(f) \; .
\end{equation}
Assuming stationary noise, the one-sided spectral density $S_h(|f|)$
is given by $\langle\tilde{h}(f)\tilde{h}(f')^*\rangle=S_h(|f|)
\delta(f-f')$, where $\langle\ldots\rangle$ denotes an expectation
over all random phases of all plane wave modes contributing to the
noise, and $^*$ denotes complex conjugation.  Taking mode amplitudes
and directions to be uncorrelated, this gives:
\begin{equation}
\label{eq:pressure-h-noise}
S_h(|f|) = \left[\frac{G\rho c}{4\pi^2\gamma Lf^3}
C(2\pi fr_{\mathrm{min}}/c) \right]^2
	\langle\cos^2\theta\rangle \frac{S_p(|f|)}{p^2} \; ,
\end{equation}
where $S_p(|f|)$ is the acoustic noise spectral density measured
outside the building in the vicinity of a particular test mass.  Since
the two test masses in an arm are many wavelengths apart, their noise
will be uncorrelated, and will thus add in noise power.  Similarly,
the noise from the two arms will add in power.  (Actually this is a
bit of an overestimate, since the noise in the motion of the test
masses at the corner station will be somewhat correlated.)  Noting
that $\langle\cos^2\theta\rangle=1/3$, one finds that the total noise
in the gravitational wave signal is:
\begin{equation}
\label{eq:pressure-noise-total}
S_h(|f|) = \left(\frac{G\rho c}{4\pi^2\gamma L}\right)^2
	\frac{1}{3f^6p^2} \sum_{i=1}^4
	C(2\pi fr^{(i)}_{\mathrm{min}}/c)^2S_p^{(i)}(|f|) \; ,
\end{equation}
where $i$ denotes a particular test mass in the interferometer,
$r^{(i)}_{\mathrm{min}}$ is the dead air radius about the $i$th test
mass, and $S_p^{(i)}(|f|)$ is the acoustic noise spectrum measured
outside the building enclosing that test mass.

\begin{figure}[t]
\epsfig{file=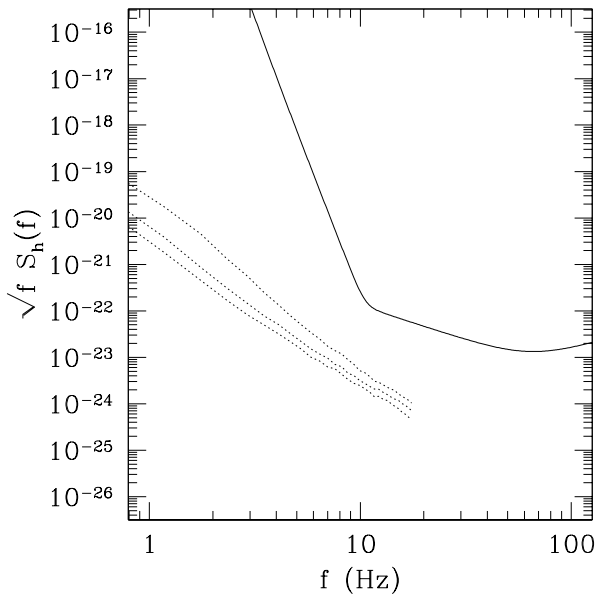,width=3.25in}
\caption{\label{fig:pressure-noise} Plot of dimensionless strain noise
$\sqrt{fS_h(f)}$ versus frequency $f$ for infrasonic atmospheric
gravity gradients.  The solid curve is the projected noise floor for
advanced LIGO detectors; the dotted curves are the first, second, and
third quartiles of noise produced by gravity gradients from ambient
pressure waves.  Data for the infrasonic noise power are taken from
Fig.~3 of~\cite{Posmentier_E:1974}.  Clearly the pressure waves would
not contribute significantly to LIGO noise in any frequency range.}
\end{figure}

Infrasound noise spectra should be taken at the actual interferometer
sites, but one can make estimates based on typical terrestrial
atmospheric noise.  An empirical study~\cite{Posmentier_E:1974}
collected 256 power spectra of 1--16~Hz infrasound data from a rural
forest 50km from New York City over a period of months, and found that
the average noise spectrum $S_p(f)$ was relatively flat at 6--16~Hz,
though with widely varying amplitude: 25\% of the spectra had noise
under $\sim100\mathrm{nbar}^2/\mathrm{Hz}$, 50\% under
$\sim300\mathrm{nbar}^2/\mathrm{Hz}$, and 75\% under
$\sim1000\mathrm{nbar}^2/\mathrm{Hz}$.  I use
Eq.~(\ref{eq:pressure-noise-total}) to compute the corresponding noise
in the LIGO detector.  Assuming that the end masses (with
$r_{\mathrm{min}}\sim5\mathrm{m}$) dominate the contribution to the
gravity gradient noise and contribute equally, the noise in the
gravitational-wave signal around 10~Hz is:
\begin{eqnarray}
S_h(|f|) & \sim & (6\times10^{-48}\mathrm{Hz}^{-1})
\left(\frac{f}{10\mathrm{Hz}}\right)^{-6}
\left(\frac{S_p(|f|)}{1000\mathrm{nbar}^2/\mathrm{Hz}}\right)
\nonumber\\
& & \times C(f/10\mathrm{Hz}) \; .
\label{eq:pressure-noise-typical}
\end{eqnarray}
The results are plotted in Fig.~\ref{fig:pressure-noise}, using
infrasound power spectra read off of Fig.~3
of~\cite{Posmentier_E:1974}.

Even the third-quartile power spectrum is between two and three orders
of magnitude below the expected noise floor of
$2\times10^{-45}\mathrm{Hz}^{-1}$ at 10~Hz projected for advanced LIGO
interferometers.  Thus ambient infrasound is probably a negligible
effect for determining the noise floor for most interferometric
gravitational-wave detectors.  Nonetheless the issue cannot be
completely resolved without infrasonic noise data from the actual
interferometer sites.

\subsection{Ground absorption}
\label{ss:ground-absorption}

In the derivation above I treated the sound waves as being reflected
off the ground.  Now consider what happens if a sound wave is absorbed
by the ground.  The energy flux in a traveling compression wave is
$C^{3/2}\rho^{-1/2}\langle(\delta x/x)^2\rangle$, where $C$ is the
compression modulus of the medium ($C_{\mathrm{air}}=\gamma
p_{\mathrm{air}}$), and $\langle(\delta x/x)^2\rangle$ is the average
squared dimensionless compression factor over a wave cycle.  The
gravity gradient noise induced by such a wave goes as
$c^2\langle(\delta\rho)^2\rangle$, where
$\langle(\delta\rho)^2\rangle\sim\rho\langle(\delta x/x)^2\rangle$,
and $c=\sqrt{C/\rho}$ is the wave speed in that medium.  Therefore, if a
sound wave is completely absorbed by the ground, the resulting ground
motions will produce gravity gradient noise contributions in the
ratio:
\begin{equation}
\label{eq:ground-air}
\frac{S_h^{\mathrm{(ground)}}}{S_h^{\mathrm{(air)}}} =
	\left(\frac{\rho_{\mathrm{ground}}}
		{\rho_{\mathrm{air}}}\right)^{3/2}
	\left(\frac{C_{\mathrm{ground}}}
		{C_{\mathrm{air}}}\right)^{-1/2} \; .
\end{equation}

Now for $C_{\mathrm{air}}=\gamma
p_{\mathrm{air}}=1.4\times10^5\mathrm{N}\,\mathrm{m}^{-2}$,
$C_{\mathrm{ground}}\sim3\times10^8\mathrm{N}\,\mathrm{m}^{-2}$,
$\rho_{\mathrm{air}}=1.3\mathrm{kg}\,\mathrm{m}^{-3}$,
$\rho_{\mathrm{ground}}\sim1.8\times10^3\mathrm{kg}\,\mathrm{m}^{-3}$,
the ratio turns out to be of order $10^3$; that is, if sound waves
were completely absorbed by the ground, the resulting ground
vibrations would produce gravity gradient noise levels about 1000
times greater than the atmospheric gravity gradients.  However, it was
shown in \cite{Hughes_S:1998} that seismic gravity gradients are only
just large enough to worry about.  So the only way that atmospheric
gravity gradients can be larger or of the same order as seismic
gravity gradients is if the waves are mostly reflected off of the
ground.  This was one of the assumptions used in deriving
Eq.~(\ref{eq:pressure-gradient-interferometer}).

\section{Atmospheric temperature perturbations}
\label{s:atmospheric-temperature-perturbations}

The largest small-scale atmospheric density perturbations are caused
not by pressure waves but by temperature perturbations.  As heat is
transported up through a convective atmospheric layer, convective
turbulence mixes pockets of warm and cool air to form temperature
perturbations on all lengthscales down to a few millimetres.  On the
timescales of interest (less than a second) these perturbations are
effectively ``frozen'' into the airmass, while pressure differences
disperse rapidly in the form of sound waves.  Perturbations in the air
density $\rho\propto p/T$ are therefore caused predominantly by the
temperature perturbations, which are typically several orders of
magnitude larger than the pressure perturbations.  Although they are
frozen into the airmass, these temperature perturbations can cause
rapid time-varying density fluctuations $\delta\rho=-\rho\delta T/T$
as the wind carries them past a point in space.  This is the primary
source of ``seeing'' noise that affects optical astronomy.

The appendix to this chapter gives a rigorous mathematical derivation
of the gravity gradient noise spectrum due to these temperature
perturbations.  This section gives a qualitative derivation that
reproduces the final result to order of magnitude.

\begin{figure*}
\epsfig{file=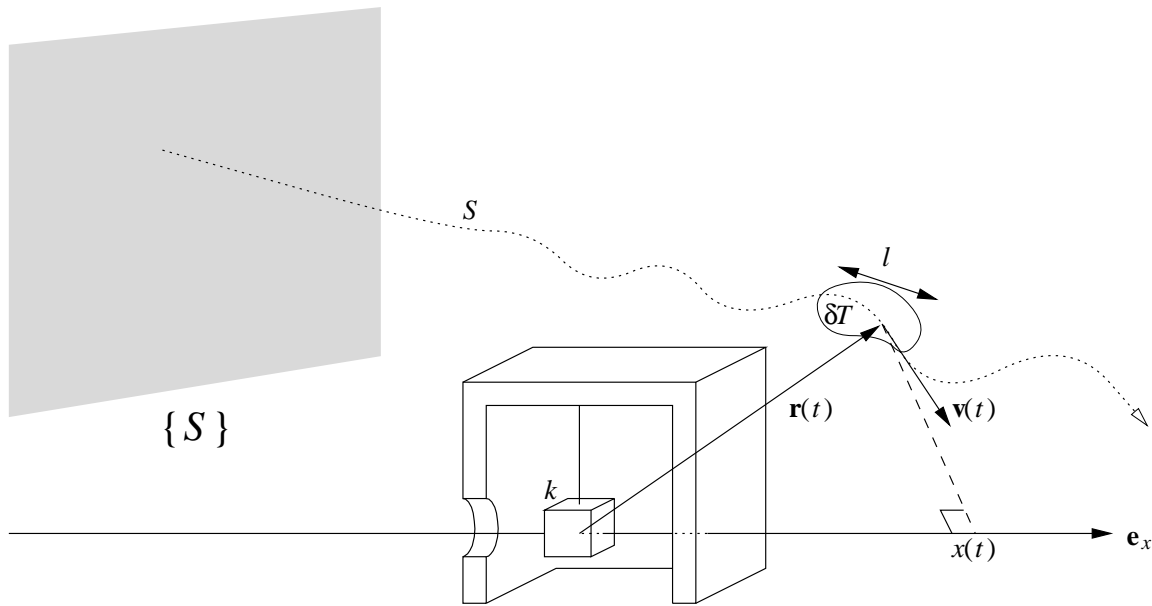,width=6in}
\caption{\label{fig:streamlines-approx} Schematic of a pocket of air
with temperature perturbation $\Delta T$ over a lengthscale $l$, being
advected past a test mass $k$ along a streamline $S$.  The pocket has
an instantaneous velocity $\mathbf{v}(t)$ and position $\mathbf{r}(t)$
relative to the test mass, and $x(t)$ is the projection of
$\mathbf{r}(t)$ onto the axis of the interferometer arm.  $\{S\}$
denotes a reference plane intersecting orthogonally with all
streamlines $S$.}
\end{figure*}

The gravity gradient signal at some frequency $f$ is caused by pockets
of warm or cool air with some lengthscale $l$ being advected past the
interferometer test mass at a speed $v$, where $l\sim v/2\pi f$.
Consider a single such pocket of air with a temperature perturbation
$\delta T$ away from the ambient temperature $T$.  The gravitational
acceleration produced in the instrument as a function of time $t$ is
$g_x(t)=G\rho l^3(\delta T/T)x(t)r^{-3}(t)$, where $\rho$ is the
ambient air density, $r(t)$ is the distance of the air pocket from the
test mass as it is blown past, and $x(t)$ is this distance projected
onto the axis of the interferometer arm.  This geometry is sketched in
Fig.~\ref{fig:streamlines-approx}.

Now in general, the noise power spectral density in any quantity $a$
due to a background of independent, uncorrelated events is
$S_a(|f|)=(2/\Delta t)|\tilde{a}(f)|^2$, where $\tilde{a}(f)$ is the
Fourier spectrum from a single event and $\Delta t$ is the spacing
between events.  Assuming uncorrelated pockets of air, then
independent pockets of air arrive along any given streamline at
intervals $\Delta t\sim l/v$, and streamlines separated by more than
$l$ add noise incoherently (i.e., add linearly in power).  This gives
the following noise spectrum:
\begin{equation}
\label{eq:g-noise-approx}
S_g(|f|) \sim \frac{2l}{v} \int_{\{S\}}\frac{dA}{l^2}
	\left(\frac{G\rho}{T}\right)^2 \delta T^2(l)
	|\tilde{G}_S(f)|^2 \; ,
\end{equation}
where $\int_{\{S\}}$ denotes an integral over a plane crossing all
streamlines $S$, and $\tilde{G}_S(f)$ is the Fourier transform of the
function $G(t)=x(t)/r^3(t)$ taken along a given streamline.  The
quantity $\delta T^2(l)$ is the average squared temperature difference
between points a distance $l$ apart.  Turbulent mixing theory, as well
as actual micrometeorological measurements, predict a power-law
behaviour for small separations: $\delta T^2(l)\sim c_T^2l^p$, where
$p$ is typically $2/3$.  This applies for horizontal separations $l$
up to of order 50~times the height of a given air pocket above the
ground~\cite{Busch_N:1972}.  For streamlines more than a metre or so
above ground level, then, this behaviour for $\delta T^2$ should be
good out to distances $l\sim50$m, corresponding to frequencies
$\agt0.2\mathrm{Hz}(v/10\mathrm{m}\,\mathrm{s}^{-1})$.

Eq.~(\ref{eq:g-noise-approx}) gives the gravity gradient noise on a
given test mass in the interferometer.  Denoting the test masses by
the index $k=1\ldots4$, and assuming that each test mass contributes
independently to the noise in the gravitational-wave signal $h$, one
has $S_h(|f|)=(2\pi f)^{-4}L^{-2}\sum_{k}S_g(|f|)$.  Combining this
with Eq.~(\ref{eq:g-noise-approx}) and the relation $l\sim v/2\pi f$
yields:
\begin{eqnarray}
S_h(|f|) & \sim & 2\left(\frac{G\rho}{LT}\right)^2 c_T^2
	(2\pi f)^{-(p+7)} v^{p+4} \nonumber\\
& & \times\sum_k \int_{\{S\}} dA |\tilde{G}_{S,k}(f)|^2 \; .
\label{eq:h-noise-approx}
\end{eqnarray}

The more rigorous analysis in the appendix
(Sec.~\ref{a:temperature-noise-spectrum}) gives an expression with
roughly the same form, but covers the factors of order unity, and also
accounts for the fact that wind speed can vary along a streamline and
between streamlines.  The more accurate formula is:
\begin{eqnarray}
S_h(|f|) & = & 2\pi^2\left(\frac{G\rho}{LT}\right)^2 c_T^2
	(2\pi f)^{-(p+7)} \sin(p\pi/2)\Gamma(p+2) \nonumber\\
& & \times\sum_k \int_{\{S\}}
	\tilde{F}_{S,k}(f)^*\tilde{G}_{S,k}(f) w\,dA \; ,
\label{eq:h-noise}
\end{eqnarray}
where $w$ is the wind speed of the streamline as it crosses the plane
of integration, and $\tilde{F}_{S,k}(f)$ and $\tilde{G}_{S,k}(f)$ are
Fourier transforms of functions $F_{S,k}(t)$ and $G_{S,k}(t)$
describing the motion of a point along a streamline $S$ past a test
mass $k$, of the form:
\begin{eqnarray}
\label{eq:streamline-moment-1}
F(t) & = & \frac{x(t)}{r(t)^3}v(t)^{p+3} \; , \\
\label{eq:streamline-moment-2}
G(t) & = & \frac{x(t)}{r(t)^3} \; .
\end{eqnarray}

It is worth noting that the frequency structure of $S_h(|f|)$ depends
on the time behaviour of the functions $x(t)$ and $r(t)$ describing
the position of a point on a streamline relative to a test mass, which
can be some distance away.  The minimum distance $r_{\mathrm{min}}$
from the test mass to the passing air is thus an additional important
scale in the problem: if $x(t)$ and $r(t)$ change significantly only
on timescales $\sim r_{\mathrm{min}}/v$, then the noise spectrum will
be cut off at frequencies $\agt v/2\pi r_{\mathrm{min}}$.

By comparison, the \emph{temperature} noise spectrum $S_T(|f|)$
measured at a point depends only on the local properties of the
atmosphere at that point.  Applying similar order-of-magnitude
arguments, one can write $S_T(|f|)\sim
(2l/v)c_T^2l^p|\tilde{H}(f)|^2$, where $H(t)\sim1$ for $|t|\alt l/v$
and 0 otherwise.  At high frequencies
$\agt0.2\mathrm{Hz}(v/10\mathrm{m}\,\mathrm{s}^{-1})$ the system
involves only one lengthscale $l$, giving the spectrum a power-law
dependence:
\begin{equation}
\label{eq:t-noise-approx}
S_T(|f|) \sim 2c_T^2 v^p (2\pi f)^{-(p+1)} \; .
\end{equation}
This is the same, to order of magnitude, as the exact result given in
Eq.~(\ref{eq:t-noise}).

\subsection{Uniform airflow}

\begin{figure*}
\epsfig{file=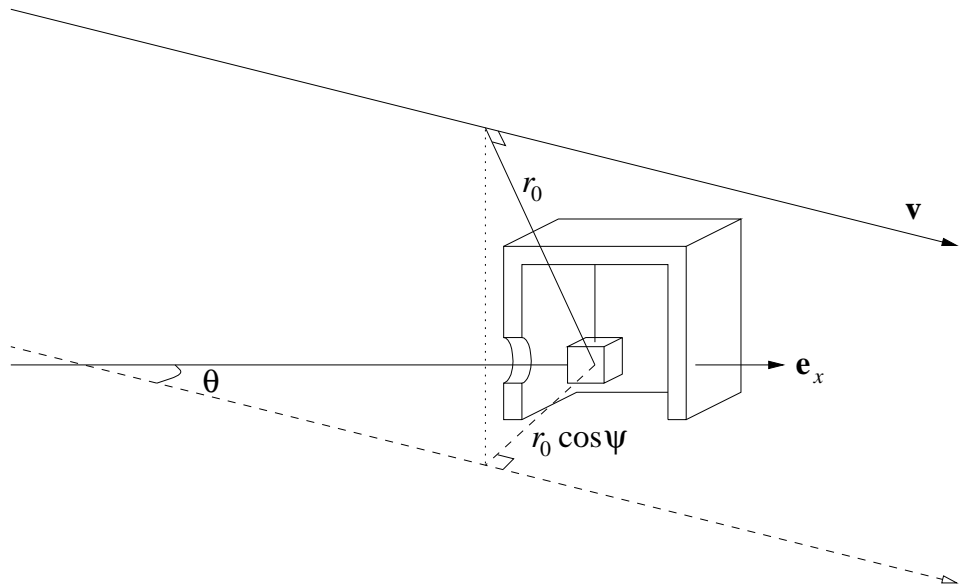,width=5in}
\caption{\label{fig:uniform-flow} Schematic of a uniform airflow
streamline passing an interferometer corner station.  $\theta$ is the
angle between the airflow and the axis of the interferometer arm,
$r_0$ is the distance from the test mass to the streamline at closest
approach, $r_0\cos\psi$ is the projection of this distance onto the
ground, and $\mathbf{v}=$constant is the velocity of the airflow.}
\end{figure*}

The gravity gradient noise is easy to compute from
Eq.~(\ref{eq:h-noise}) for the case of uniform airflow parallel to the
ground with some constant velocity $\mathbf{v}$.  Placing the
reference plane orthogonal to $\mathbf{v}$ and passing through the
test mass, the equations of motion for a given streamline past the
test mass become quite simple: $v(t)=w=v$,
$x(t)=vt\cos\theta+r_0\sin\theta\cos\psi$,
$r(t)=(r_0^2+v^2t^2)^{1/2}$, where $r_0$ is the distance from the test
mass to the nearest point on the streamline, and $r_0\cos\psi$ is the
projection of this distance onto the ground.  The geometry of the
airflow is shown in Fig.~\ref{fig:uniform-flow}.  It is easy to show
that:
\begin{eqnarray}
\tilde{G}(f) & = & -\frac{4\pi f}{v^2}
		\left[i\cos\theta\,K_0(2\pi fr_0/v)\right. \nonumber\\
\label{eq:uniform-streamline-moment-1}
& & \hspace{3em}
	\left. +\sin\theta\cos\psi\,K_1(2\pi fr_0/v)\right] \; , \\
\label{eq:uniform-streamline-moment-2}
\tilde{F}(f) & = & v^{p+3}\tilde{G}(f) \; ,
\end{eqnarray}
where $K_0$ and $K_1$ are moddified Bessel functions of the second
kind of orders 0 and 1.  I perform the integral over the above-ground
half of the reference plane, out from some radius $r_{\mathrm{min}}$
that is roughly the closest distance that the outside air can approach
the test mass.  This gives a noise contribution from a single test
mass equal to:
\begin{eqnarray}
S_h(|f|) & = & 8\pi^2\cos(\pi[p-1]/2)\Gamma(p+2)
		\left(\frac{G\rho r_{\mathrm{min}}}{TL}\right)^2
	\nonumber\\
& \times & c_T^2 \left(\frac{v}{2\pi f}\right)^p (2\pi f)^{-5}
	\bigg\{\cos^2\theta\left[K_0^2(x)-K_1^2(x)\right]
	\nonumber\\
& & + \frac{1}{2}\sin^2\theta
	\left[K_1^2(x)-K_0(x)K_2(x)\right]\bigg\} \; ,
\label{eq:uniform-streamline-noise}
\end{eqnarray}
where $x=2\pi fr_{\mathrm{min}}/v$.

For typical values $v\sim10$m/s and $r_{\mathrm{min}}\sim5$m, at
frequencies above 10~Hz or so, one has $x\sim30$ or more, well into
the exponentially damped regime of the Bessel functions.  Even for
gale-force winds of 30m/s or so, the argument $x$ of the Bessel
functions will still be of order 10 or more.  The asymptotic
expansions of $K_0$, $K_1$, and $K_2$ give $K_0^2(x)-K_1^2(x)\sim
K_1^2(x)-K_0(x)K_1(x)\sim\pi e^{-2x}/(2x^2)$.  Also, I note that the
total noise will be dominated by the contribution from the two end
stations, which have smaller $r_{\mathrm{min}}$ than the corner
station.  So the total noise in the interferometer for uniform airflow
is:
\begin{eqnarray}
S_h(|f|) & = & (2\pi)^3\cos(\pi[p-1]/2)\Gamma(p+2)
	\left(\frac{G\rho}{TL}\right)^2 c_T^2 \nonumber\\
& & \times\left(\frac{v}{2\pi f}\right)^p
	(2\pi f)^{-5} e^{-4\pi fr_{\mathrm{min}}/v} \; .
\label{eq:asymptotic-noise}
\end{eqnarray}
I consider ``typical'' values of $p=2/3$,
$\rho=1.3\mathrm{kg}\,\mathrm{m}^{-3}$, $T=300\mathrm{K}$,
$L=4000\mathrm{m}$, and $r_{\mathrm{min}}=5\mathrm{m}$.  The
parameters $v$ and $c_T^2$ can vary on a minute-by-minute basis, and
should really be measured at the site of a given interferometer.
However, $c_T^2\sim0.2\mathrm{K}^2\,\mathrm{m}^{-2/3}$ is a typical
daytime peak temperature fluctuation index~\cite{Coulman_C:1985}, and
$v\sim20\mathrm{m/s}$ might be typical of a fairly windy day.  At
frequencies around 10~Hz, this gives a noise spectrum of:
\begin{equation}
\label{eq:typical-uniform-noise}
S_h(|f|) \sim (1.6\times10^{-40}\mathrm{Hz}^{-1})
	\left(\frac{f}{10\mathrm{Hz}}\right)^{-23/3}
	10^{-14(f/10\mathrm{Hz})} \; .
\end{equation}
The two dotted lines in Fig.~\ref{fig:temperature} show the gravity
gradient noise spectra computed from
Eq.~(\ref{eq:typical-uniform-noise}) for wind speeds of 10m/s and
30m/s.

Clearly the exponential cutoff makes this a negligible source of noise
for LIGO or similar detectors.  Physically this cutoff arises from the
fact that the gravity from a particular temperature perturbation
passing near the end station will affect the test mass coherently over
the second or so that it takes to travel the width of the end station.
Thus, even though the temperature noise spectrum has a high-frequency
power law tail (reflecting the fact that temperature perturbations
exist on all lengthscales), the gravity gradient signal will have this
much sharper exponentially cut-off tail.

\subsection{Potential flow near the end station}

As described above, uniform airflow is not likely to produce much
atmospheric gravity gradient noise in the pass-band of interferometric
detectors, since the shortest timescale over which the gradients
change is of order the wind crossing time of the interferometer
buildings.  However, if an air pocket could be made to accelerate over
shorter timescales, it might produce a stronger gravity gradient
signal at high frequencies.

\begin{figure}[t]
\epsfig{file=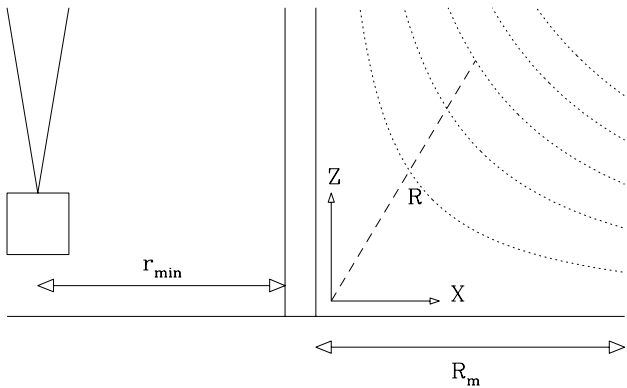,width=3.25in}
\caption{\label{fig:corner-flow} Schematic of the airflow streamlines
around the corner of an end station, assuming an incompressible
vorticity-free flow.  The dotted lines are the streamlines,
$r_{\mathrm{min}}$ is the minimum distance that the airflow can
approach the test mass, $R_m$ is the scale distance at which the flow
velocity $v$ approaches the free-streaming speed $V$, and $X$, $Z$,
and $R$ are coordinates used to describe the velocity field.}
\end{figure}

One possibility is the acceleration of the air as it is forced up and
around the wall of an end station: streamlines that approach the
ground-wall corner of the end station can have curvature scales much
shorter than the building size.  Treating the flow as incompressible
and vorticity-free, the resulting velocity field near the corner
is (p.~27 of~\cite{Landau_L:1989}):
\begin{equation}
\label{eq:corner-velocity-field}
v_x = -2AX \; , \quad v_z = 2AZ \; ,
\end{equation}
where $A=v/2R$ is a constant, and $R$, $X$, and $Z$ are measured from
the corner.  This approximation is clearly only good near the corner,
since it gives velocity increasing monotonically with radius; one
would expect $v$ to approach the free-streaming airspeed $V$ at a
distances $R_m$ of order the building height.
Fig.~\ref{fig:corner-flow} shows a schematic of the airflow near the
corner.

It is clear from Eq.~(\ref{eq:corner-velocity-field}) that, although
the streamlines are sharply curved near the corner of the end station,
the advection speed is smaller in direct proportion.  The shortest
timescale over which the motion can change is of order $R_m/V$ for all
streamlines.  Thus the streamlines close to the corner will contribute
no more high-frequency noise than the streamlines further out, at
distances of order $R_m$, and the spectrum should not differ greatly
from the one for uniform flow,
Eqs.~(\ref{eq:asymptotic-noise}),~(\ref{eq:typical-uniform-noise}).

\subsection{Vortices}
\label{ss:vortices}

Perhaps the most serious contenders for high-frequency atmospheric
gravity gradient noise are circulating vortices of air near the end
stations.  This is somewhat stretching the assumptions of the
formalism I have established, since I had previously separated the
effects of the homogeneous turbulence (which establishes the
temperature perturbations on various lengthscales), and wind flow
(which carries these perturbations past the instrument).  However, the
results in this section are only expected to be good to order of
magnitude anyway, in the absence of detailed hydrodynamic analysis of
airflow past a particular interferometer building.  I therefore apply
the above formalism to a simple model for a turbulent-like flow past
an interferometer end station.

\begin{figure}[t]
\epsfig{file=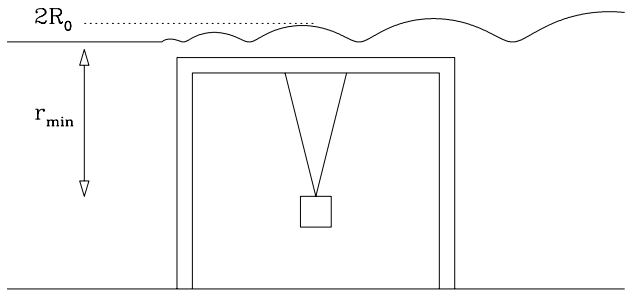,width=3.25in}
\caption{\label{fig:turbulent-flow} Schematic of a vortex-like
streamlines across the top of an end station.  The quantity
$r_{\mathrm{min}}$ is the minimum distance that the airflow can
approach the test mass, and $R_0$ is the typical radius of circulation
at this distance.}
\end{figure}

For simplicity, I specialize to airflow along the axis of an
interferometer arm, since these give the largest gravitational
accelerations.  A simple model for the turbulent flow is to take a
uniform flow and then add a cycloidal motion to it.  This gives
$x(t)\sim vt-R\sin(vt/R)$, $z(t)\sim r_0-R\cos(vt/R)$,
$r(t)=\sqrt{z^2(t)+x^2(t)}$, where $R$ is the radius of the cycloidal
motion, and $r_0\sim r_{\mathrm{min}}$ is the distance from the test
mass to the unperturbed streamline.  I treat the speed $v$ along the
streamline as a constant.  If $R$ is also a constant, then one would
expect the Fourier transform of $G(t)=x(t)/r^3(t)$
[Eq.~(\ref{eq:streamline-moment-2})] to have a spike at frequency
$f=v/2\pi R$, with a width of order $\sim v/r_0$.  However, to give a
somewhat more realistic behaviour for $R$, I treat it as growing from
zero at the leading edge of the end station to some scale value $R_0$
over the half-length $\sim r_{\mathrm{min}}$ of the end station:
$R(t)=R_0\sqrt{vt/r_{\mathrm{min}}}$.  The square-root dependence
mimics the less-than-linear growth of the thickness of a boundary
layer.  Although quite crude, this model covers the essential features
of a turbulent flow past the building: a uniform translation,
accompanied by circulating motions over a range of radii with a scale
set by $R_0$, with both the uniform flow rate and the circulation
speed set by the free-streaming airspeed $v$.  A typical streamline of
this type is shown in Fig.~\ref{fig:turbulent-flow}.

The Fourier transform of $G(t)$ is too complicated to perform
analytically, but is simple enough to compute numerically using a fast
Fourier transform.  The resulting $\tilde{G}(f)$ has the usual
exponential cutoff with frequency scale $v/2\pi r_{\mathrm{min}}$, as
for a smooth streamline, but then rises to a second peak value of
$\sim(4/v)\sqrt{R_0^3/r_{\mathrm{min}}^5}$ at a frequency
$f\sim0.06v/R_0$ and decreases from there as $f^{-3}$.  This
high-frequency tail is the power law that one would expect from the
cusps on the bottom of the cycloid.  Since I treat $v$ as constant,
$\tilde{F}(f)=v^{p+3}\tilde{G}(f)$.  The cross-sectional area of
streamlines that contribute significantly around this peak frequency
is of order $\sim2r_{\mathrm{min}}R_0$.  Plugging these into
Eq.~(\ref{eq:h-noise}) one gets a noise spectral density of
\begin{equation}
\label{eq:vortex-h-noise}
S_{h(\mathrm{max})} \sim 1.5\times10^6\left(\frac{G\rho}{LT}\right)^2
	c_T^2 R_{\mathrm{max}}^{35/3} r_{\mathrm{min}}^{-4} v^{-5}
\end{equation}
at the peak frequency of $f\sim0.06v/R_0$, where I have assumed
$p=2/3$.

\begin{figure}[t]
\epsfig{file=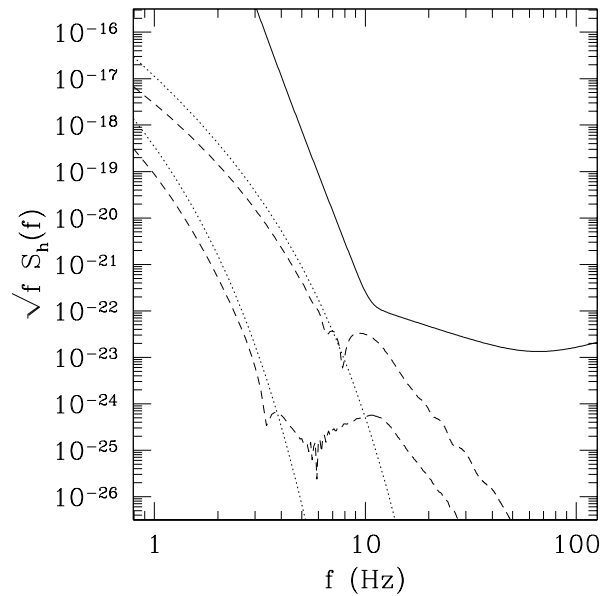,width=3.25in}
\caption{\label{fig:temperature} Plot of dimensionless noise amplitude
$\sqrt{fS_h(f)}$ versus frequency $f$.  The solid line is the
projected noise floor for advanced LIGO detectors.  The two dotted
lines are the gravity gradient noise levels caused by temperature
perturbations advected along smooth streamlines at 10m/s (left curve)
and 30m/s (right curve).  The dashed lines are the noise levels caused
by temperature perturbations advected along cycloidal vortices, as
described in Sec.~\ref{ss:vortices}, at 10m/s (left curve) and 30m/s
(right curve); in each case the size scales $R_0$ of the vortices have
been tuned to maximize the noise at 10~Hz.  Even given these
fine-tunings, the gravity gradient noise curve in the worst-case
scenario is still nearly an order of magnitude (in amplitude) below
the advanced LIGO sensitivity.}
\end{figure}

For wind speeds around 10m/s, a cycloid radius
$R_0\sim0.06\mathrm{m}(v/10\mathrm{m/s})$ puts the noise peak at the
10~Hz seismic noise wall for advanced LIGO detectors.  The atmospheric
gravitational noise contribution from a single end station is then:
\begin{eqnarray}
S_{h(\mathrm{max})} & \sim & (1.3\times10^{-49}\mathrm{Hz}^{-1})
	\left(\frac{c_T^2}{0.2\mathrm{K m}^{-2/3}}\right) \nonumber\\
& & \times\left(\frac{v}{10\mathrm{m/s}}\right)^{20/3}
	\left(\frac{r_{\mathrm{min}}}{5\mathrm{m}}\right)^{-4} \; .
\label{eq:typical-vortex-noise}
\end{eqnarray}
This is over five orders of magnitude below the expected advanced LIGO
noise floor of $2\times10^{-45}\mathrm{Hz}^{-1}$ at 10~Hz.  Gale-force
winds ($v\sim30$m/s) will bring this up to
$2\times10^{-46}\mathrm{Hz}^{-1}$, still an order of magnitude below
the advanced LIGO noise curve.  The dashed lines in
Fig.~\ref{fig:temperature} show the actual data from the numerical
Fourier transforms for these two cases.

Since even the worst-case estimate is still an order of magnitude
below the advanced LIGO noise floor (as well as the seismic gravity
gradient noise floor in~\cite{Hughes_S:1998}), it seems unlikely that
turbulent vortices will be sufficient to raise the atmospheric
gravitational noise to significant levels, even given the
approximations made in this analysis.  However, to settle this matter
definitively would require a much more sophisticated numerical
analysis of the temperature perturbations and airflow past the
buildings of a particular facility.

\section{Shockwaves}
\label{s:shockwaves}

Although atmospheric pressure waves are unlikely to be a significant
source of gravity gradient noise in interferometric gravitational-wave
detectors, the sudden pressure changes caused by atmospheric
shockwaves could potentially produce detectable transient signals in
the detector, if such shocks occur in the vicinity of the detectors.
Shocks are specifically a matter of concern because they can produce
significant pressure changes over timescales less than 0.1~s,
corresponding to the lower end of the pass-bands of most
interferometric detectors.

Consider a shock that produces a sudden jump in air pressure in the
vicinity of one of the interferometer test masses: $\delta p(t)=\Delta
p\Theta(t-t_0)$.  It is a simple matter to take the Fourier transform
and apply Eq.~(\ref{eq:pressure-h-tilde}) to obtain:
\begin{equation}
\label{eq:shock-h-tilde}
\tilde{h}(f) = \frac{G\rho c}{8\pi^3\gamma Lf^4} \frac{\Delta p}{p}
	ie^{2\pi ft_0} \cos(\theta) C(2\pi fr_{\mathrm{min}}/c) \; .
\end{equation}
Here, $\theta$ is the angle between the interferometer arm and the
normal to the shock front.  If the shock has a finite rise time
$\Delta t$, we can mimic this analytically by convolving in the time
domain with a Gaussian of width $\sigma\sim\Delta t$.  In the
frequency domain this multiplies our amplitude by a Gaussian of width
$\sigma\sim1/\Delta t$, giving an exponential cutoff at frequencies
above $1/\Delta t$.  Typical shocks from, for instance, supersonic
objects have rise times on the order of a few
milliseconds~\cite{Hayes_W:1971}, corresponding to cutoff frequencies
of a few hundred~Hz.  However, one expects the dominant contribution
of the signal to come from much lower frequencies, before the
building-size cutoff factor $C$ kicks in.

Shocks are transient phenomena that will produce \emph{signals} in the
detector, rather than raising the noise floor.  What one would like to
know is what signal-to-noise ratio the shock will produce.  In general
this depends on what filters one is using to search for signals, and
how well these filters overlap with the signal produced by a shock.
However, the signal from a shock is likely to overlap quite well with
templates designed to search for generic impulsive phenomena; such
templates are likely to be used in advanced interferometers as control
over nonstationary instrumental noise improves.  Thus it is reasonable
to consider the signal-to-noise ratio $\rho$ that a matched filter
would give~\cite{Thorne_K:1999}:
\begin{equation}
\label{eq:snr}
\rho^2 	= \int_0^\infty \frac{4\tilde{h}(f)\tilde{h}^*(f)}{S_h(f)}\,df
	= \int_{-\infty}^\infty \frac{|2f\tilde{h}(f)|^2}{fS_h(f)}
		\,d\ln f \; .
\end{equation}
This shows, roughly speaking, that the relative magnitude of the
dimensionless signal amplitude $2|f\tilde{h}(f)|$ and the
dimensionless noise amplitude $\sqrt{fS_h(f)}$ over a logarithmic
frequency interval gives a good indication of the signal-to-noise
ratio produced in the detector.

\subsection{Sonic booms}
\label{ss:sonic-booms}

Sonic booms caused by supersonic bodies are one example of atmospheric
shocks that might affect interferometric gravitational-wave detectors.
Direct shockwaves from a supersonic aircraft will typically hit the
ground in a ``carpet'' about 15--20~km wide under the aircraft's
flight path.  Outside this carpet, the temperature gradient near the
ground will completely reflect the shockwave before it touches down.
However, shockwaves will also reflect downward off of the temperature
inversion in the stratosphere and thermosphere, forming secondary and
higher-order ``carpets'' out to many hundreds of
kilometres~\cite{Balachandran_N:1977}.  The presence or absence of
these higher-order waves can depend quite sensitively on conditions in
the upper atmosphere.

A detailed prediction of these effects is beyond the scope of this
paper.  However, to give an indication of their potential seriousness,
I will consider what would happen if a supersonic aircraft were
actually to overfly the instrument at a height of several kilometres.

The characteristic profile of a sonic boom is a symmetric N-wave,
consisting of a shock that increases the pressure by an amount $\Delta
p$, followed by a smooth decrease in pressure of $2\Delta p$ over a
time $\Delta t$, followed by a second rising shock $\Delta p$ to
restore the ambient pressure.  According to Eq.~(9.78)
of~\cite{Witham_G:1974}, the strength of the shocks is:
\begin{equation}
\label{eq:boom-dp}
\frac{\Delta p}{p} \approx \frac{2^{1/4}\gamma}{(\gamma+1)^{1/2}}
	(M^2-1)^{1/8} \kappa\delta l^{3/4}r^{-3/4} \; ,
\end{equation}
where $\gamma$ is the adiabatic coefficient of air (1.4 at normal
temperatures), $M$ is the Mach number of the aircraft (its speed
divided by the sound speed), $\kappa$ is a dimensionless form factor
that depends on the shape of the aircraft (typically around 1), $l$ is
the length of the aircraft, $\delta$ is the ratio of the aircraft's
typical thickness to its length, and $r$ is the closest distance that
the aircraft came to the point of measurement.  Between the two
shocks, the rate of change of pressure in a direction $\mathbf{e}_x$
parallel to the line of flight is given, in Eq.~(9.80)
of~\cite{Witham_G:1974}, as:
\begin{equation}
\label{eq:boom-dpdx}
\frac{1}{p}\frac{dp}{dx} \approx \frac{\gamma}{\gamma+1}
	\frac{(M^2-1)^{1/2}}{M^2} \frac{1}{r} \; .
\end{equation}
The shock fronts move outward at the sound speed $c$ in the direction
orthogonal to their surface, while the entire cone travels in the
$\mathbf{e}_x$ direction along with the aircraft at a speed $Mc$.  The
total change of pressure between the two shocks is $2\Delta p$.  From
these facts and Eqs.~(\ref{eq:boom-dp}) and~(\ref{eq:boom-dpdx}), one
can show that the time between the two shocks is:
\begin{equation}
\label{eq:boom-dt}
\Delta t \approx 2^{5/4}(\gamma+1)^{1/2} \frac{M}{(M^2-1)^{3/8}}
	\kappa\delta \frac{l^{3/4}r^{1/4}}{c} \; .
\end{equation}
On frequency scales higher than $1/\Delta t$ (typically a few Hz for a
supersonic aircraft a few kilometres away), the sonic boom looks like
simple Heaviside shocks, giving $\delta\tilde{p}(f)\sim1/f$.  At lower
frequencies, though, the entire N-wave looks like the derivative of a
$\delta$-distribution, giving $\delta\tilde{p}(f)\sim f$.  Performing
the Fourier transform analytically and plugging into
Eq.~(\ref{eq:pressure-h-tilde}), one obtains:
\begin{eqnarray}
\tilde{h}(f) & = & \frac{G\rho c}{4\pi^3\gamma L} \frac{1}{f^4}
	\frac{\Delta p}{p} \cos(\theta) C(2\pi fr_{\mathrm{min}}/c)
\nonumber\\
& & \times\left[\frac{\sin(\pi f\Delta t)}{\pi f\Delta t} -
		\cos(\pi f\Delta t)\right] e^{2\pi ift_0} \; ,
\label{eq:boom-h-tilde}
\end{eqnarray}
where $t_0$ is the time when the midpoint of the N-wave crosses the
detector.  As expected, the amplitude goes roughly as $f^{-4}$, except
for frequencies less than $1/\Delta t$, where it goes as $f^{-1}$.

Now let us plug in some typical numbers.  The numbers
$G=6.67\times10^{-11}\mathrm{m}^3\mathrm{kg}^{-1}\mathrm{s}^{-2}$,
$\rho=1.3\mathrm{kg}\,\mathrm{m}^{-3}$, $\gamma=1.4$,
$c=332\mathrm{m}\,\mathrm{s}^{-1}$, and $L=4000\mathrm{m}$ can be
treated as constant.  A supersonic jet aircraft might have a length of
$l=10$m, a typical diameter of $\delta l=2$m, and be traveling at
Mach~$M=1.5$ at a distance of $r=10$km or so.  Let $\cos\theta$ be~1
for an upper limit.  Then $\Delta t\sim0.2$s, and for frequencies
$f\agt10$Hz the dimensionless signal amplitude is:
\begin{eqnarray}
2|f\tilde{h}(f)| & \sim & 1.4\times10^{-19} (M^2-1)^{1/8}
	C(2\pi fr_{\mathrm{min}}/c)
	\left(\frac{\delta}{0.2}\right) \nonumber\\
& & \times\left(\frac{l}{10\mathrm{m}}\right)^{3/4}
	\left(\frac{r}{10\mathrm{km}}\right)^{-3/4}
	\left(\frac{f}{10\mathrm{Hz}}\right)^{-3} \; .
\label{eq:jet-amplitude}
\end{eqnarray}
This is three orders of magnitude above the expected noise floor of
$\sqrt{fS_h(f)}\sim1.4\times10^{-22}$ at 10~Hz for advanced LIGO
interferometers!

By contrast, consider a .30-calibre rifle bullet ($l\approx0.025$m,
$\delta\approx0.3$) passing at Mach~3 within 10m of an interferometer
test mass.  (This stretches the assumption of a plane-wave shock front
at the test mass, but the order of magnitude should be correct.)  The
bullet produces a much stronger double shock, but with a time interval
$\Delta t\sim0.5$ms, so $1/\Delta t\sim2$kHz.  The low-frequency tail
of this signal will have dimensionless amplitude:
\begin{eqnarray}
2|f\tilde{h}(f)| & \sim & 1.8\times10^{-23}
	\frac{M^2(M^2-1)^{-5/8}}{2.5}C(2\pi fr_{\mathrm{min}}/c)
\nonumber\\
& \times & \left(\frac{\delta}{0.3}\right)^3
	\left(\frac{l}{0.025\mathrm{m}}\right)^{9/4}
\nonumber\\
& & \times\left(\frac{r}{10\mathrm{m}}\right)^{-1/4}
	\left(\frac{f}{10\mathrm{Hz}}\right)^{-1} \; .
\label{eq:bullet-amplitude}
\end{eqnarray}
This is nearly an order of magnitude \emph{below} the dimensionless
noise amplitude in advanced LIGO, and therefore too small to be of any
serious concern.

\begin{figure}[t]
\epsfig{file=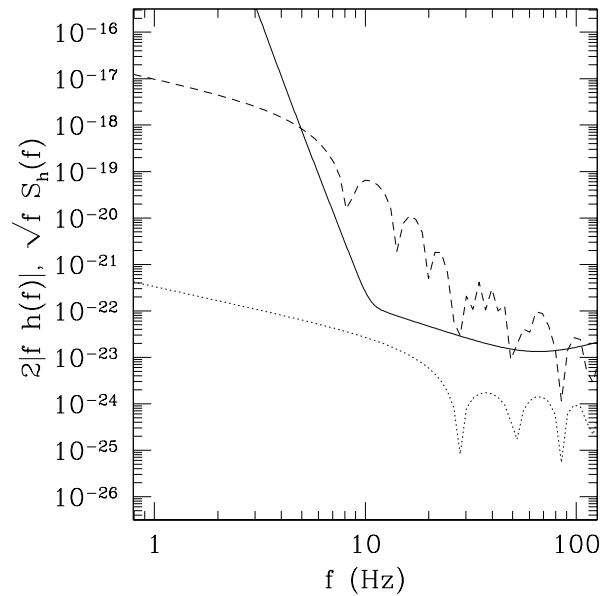,width=3.25in}
\caption{\label{fig:sonic-boom} Plot comparing dimensionless signal
strengths $2|f\tilde{h}(f)|$ of gravity gradients from sonic booms
with the dimensionless interferometer noise amplitude
$\sqrt{fS_h(f)}$, as a function of frequency.  The solid line is the
projected noise amplitude for advanced LIGO detectors.  The dotted
line is the signal from a .30~calibre bullet traveling at Mach~3
within 10m of the end station; the dashed line is the signal from a
10m long aircraft traveling at Mach~1.5 at a distance of 10km.  The
aircraft's sonic boom can produce a detectable gravity gradient
signal.}
\end{figure}

Fig.~\ref{fig:sonic-boom} shows more complete gravity gradient signal
spectra computed using Eq.~(\ref{eq:boom-h-tilde}) with the above
parameters for a supersonic aircraft and rifle bullet.  These are
plotted along with the anticipated dimensionless noise amplitude for
advanced LIGO detectors.

\subsection{Vetoing shockwave signals}

While atmospheric shockwaves are a potential source of spurious
signals in gravitational-wave detectors, they are easy to veto using
environmental sensors.  One need simply place infrasound microphones
outside the buildings and test-mass vacuum enclosures.  If these
sensors detect a pressure change of more than a millibar over
timescales of 50--100~milliseconds, then one might expect spurious
signals with dimensionless amplitude of $\sim10^{-22}$ in the
10--20~Hz frequency range.  The stretch of data containing the
potential spurion can then be discarded.

Alternatively, if the same shock profile is detected in an array of at
least three sound sensors, then one can determine the direction of
propagation of the shock and predict the actual induced test-mass
motions.  The spurious signal could then be subtracted out of the data
stream.  This is a much trickier procedure, and would only be
necessary in the unlikely event that significant amounts of data were
being corrupted.

In either case, it is clear that infrasound sensors will be important
environmental monitors for advanced interferometric detectors.

\section{High-speed objects}
\label{s:high-speed-objects}

Another potential source of spurious signals in the interferometer is
the gravity gradient caused by the motion of an individual massive
object past the interferometer end station, or the collision of such
an object with the end station.  The latter is particularly serious,
since the sudden deceleration of the object can produce a signal at
high frequencies.  The issue of human-generated gravity gradient noise
has been addressed in~\cite{Thorne_K:1999}, but there are other
sources outside the facility that must be considered, such as stray
bullets and wind-borne debris.  In particular, the Hanford LIGO
facility is plagued by tumbleweeds, which can produce non-negligible
gravity gradient signals.

The general formula for the spurious gravitational-wave signal
produced by a moving object is:
\begin{equation}
\label{eq:object-h-tilde}
\tilde{h}(f) = \frac{GM}{4\pi^2 Lf^2} \int_{-\infty}^\infty
	\frac{x(t)}{r^3(t)}e^{2\pi ift} \,dt \; ,
\end{equation}
where $M$ is the mass of the object, $r(t)$ is its distance from the
test mass as a function of time, and $x(t)$ is its distance from the
end mass in the direction parallel to the interferometer arm.  For an
object traveling parallel to the ground in a straight line at speed
$v$, Eq.~(\ref{eq:object-h-tilde}) becomes [as in
Eq.~(\ref{eq:uniform-streamline-moment-1})]:
\begin{eqnarray}
\tilde{h}(f) & = & \frac{GM}{Lv^2\pi f}
	\left[K_0\left(\frac{2\pi fr_{\mathrm{min}}}{v}\right)
		\cos\theta\right. \nonumber\\
& & \quad\left.
	- iK_1\left(\frac{2\pi fr_{\mathrm{min}}}{v}\right)
		\sin\theta\cos\psi\right] e^{2\pi ift_0} \; ,
\label{eq:passing-h-tilde}
\end{eqnarray}
where $\theta$ is the angle between the line of motion and the
interferometer arm, $r_{\mathrm{min}}$ is the distance of closest
approach between the object and the test mass, $\psi$ is the angle
projecting this distance onto the ground, and $t_0$ is the time of
closest approach.  $K_0$ and $K_1$ are modified Bessel functions of
the second kind of order 0 and 1, respectively.

Under moderately windy conditions (wind speeds up to 15m/s or so),
tumbleweeds at the Hanford LIGO facility will bounce along the ground
at 5--10m/s.  In stronger winds, the tumbleweeds become airborne, with
speeds approaching the wind speed; they can fly right over the LIGO
buildings, or impact with considerable force.  The same may be true of
wind-borne debris at other interferometer facilities.  However, the
value of $r_{\mathrm{min}}$ is usually at least 5m, so for frequencies
above 10~Hz one has $2\pi fr_{\mathrm{min}}/v\agt10$ even for very
strong winds ($v\sim30$m/s).  In this regime the Bessel functions are
exponentially damped, so these objects will not produce significant
gravity gradient signals simply by blowing past the instrument.  A
rifle bullet, on the other hand, might be moving around 1000m/s,
putting us in the small-argument regime of the Bessel functions, where
$K_0(x)\sim -\ln(x)$ and $K_1(x)\sim x^{-1}$.  Taking the most
dangerous geometry $\theta=\pi/2$, $\psi=0$, and assuming a bullet
mass of around 5~grammes, this gives a dimensionless signal amplitude
near 10~Hz of:
\begin{eqnarray}
2|f\tilde{h}(f)| & \sim & 1.6\times10^{-22}
	\left(\frac{M}{5\mathrm{g}}\right)
	\left(\frac{1000\mathrm{m/s}}{v}\right) \nonumber\\
& & \times\left(\frac{5\mathrm{m}}{r_{\mathrm{min}}}\right)
	\left(\frac{10\mathrm{Hz}}{f}\right) \; .
\label{eq:passing-bullet-signal}
\end{eqnarray}
This gives a signal-to-noise ratio of about 1 at 10~Hz, a bit below
the detectable threshold.  In fact, even if one fine-tunes the bullet
speed $v$, the largest signal-to-noise ratio that one can get at 10~Hz
is about 2, for a speed of around 250m/s.  Since events with signals
less than about 5 times the noise will probably be ignored in any
case, one can conclude that objects flying past a test mass are not
likely to be serious sources spurious events.

If an object does not pass smoothly by the interferometer but instead
collides with an end station, the signal at $\sim10$~Hz can be large
even for slow-moving objects: it is the deceleration time, not the
end-station-crossing time, that sets the frequency scale of the
signal.  Suppose an object collides end-on with the end station at a
speed $v$, coming to a stop within a distance $d$ at constant
acceleration.  Let $t=0$ denote the time that the object comes to
rest.  The motion of the object is then given by:
\begin{equation}
\label{eq:collision-motion}
r(t) = x(t) = \left\{ \begin{array}{lc}
	r_{\mathrm{min}}-d-vt \quad & t \leq -2d/v \\
	r_{\mathrm{min}}+(vt)^2/4d \quad & -2d/v \leq t \leq 0 \\
	r_{\mathrm{min}} \quad & t \geq 0 \end{array} \right. \; .
\end{equation}
The Fourier transform of $x(t)/r^3(t)=1/r^2(t)$ is tricky to do
analytically, so I have relied on numerical fast Fourier transforms,
and then made approximate analytic fits to the result.  However, one
can qualitatively predict the shape of the signal in frequency space.
The function $1/r^2(t)$ starts out near zero and then slowly rises
over a timescale $r_{\mathrm{min}}/v$ to a value $r_{\mathrm{min}}^2$,
then quickly levels off at that value over a timescale $d/v$.  So on
frequency scales $\ll v/r_{\mathrm{min}}$, $1/r^2(t)$ looks like a
step function, whose Fourier transform goes as $1/f$.  On frequency
scales $\geq v/r_{\mathrm{min}}$ but $\ll v/d$, one sees the
deceleration as a cusp (discontinuous first derivative), giving a
Fourier transform that goes as $1/f^2$.  On frequency scales $\geq
v/d$, the deceleration appears smooth, but the onset of deceleration
in sudden, giving a $1/f^3$ behaviour.  The following gives a good fit
to the numerical Fourier transform:
\begin{eqnarray}
\left|\int_{-\infty}^\infty \frac{e^{2\pi ift}}{r^2(t)}\,dt\right|
& \sim & \frac{1}{vr_{\mathrm{min}}}\left[
		5.9\left(\frac{fr_{\mathrm{min}}}{v}\right) +
		14\left(\frac{fr_{\mathrm{min}}}{v}\right)^2\right.
\nonumber\\
& & \hspace{3em}\left.
	+ 59\left(\frac{f^3r_{\mathrm{min}}^2d}{v^3}\right)
	\right]^{-1} \; .
\label{eq:collision-fit}
\end{eqnarray}
The two breakpoints separating the three branches are at frequencies
$f_1=0.4v/r_{\mathrm{min}}$ and $f_2=0.24v/d$.  More precisely, since
the deceleration occurs over a well-defined time $2d/v$, the third
branch of the Fourier transform is oscillatory with nodes every $v/2d$
in frequency space; the functional fit in Eq.~(\ref{eq:collision-fit})
is an envelope containing these oscillations.

For a 5g bullet striking an end station at 1000m/s, the signal at
10~Hz is dominated by the low-frequency tail regardless of stopping
distance.  The dimensionless signal amplitude is then:
\begin{equation}
\label{eq:colliding-bullet-signal}
2|f\tilde{h}(f)|\sim3\times10^{-22}\left(\frac{M}{5\mathrm{g}}\right)
	\left(\frac{10\mathrm{Hz}}{f}\right)^2
	\left(\frac{5\mathrm{m}}{r_{\mathrm{min}}}\right)^2 \; .
\end{equation}
The signal amplitude is about the same as for a passing bullet,
although the dependence on $f$ (and $r_{\mathrm{min}}$ and $v$) is
different.  For a tumbleweed or other wind-borne object, by contrast,
the sudden deceleration can create significant high-frequency noise.
A typical tumbleweed at Hanford has a mass of $M=0.1$kg and a diameter
of 0.4m, and can be compressed by about half that amount ($d=0.2$m).
Larger weeds can be twice as large in diameter, putting their masses
in the 0.5--1kg range~\cite{Raab_F:1999}.  For moderate to high wind
speeds ($v=10$--30m/s), the signal at frequencies above 10~Hz is
dominated by the second branch of Eq.~(\ref{eq:collision-fit}), giving
a dimensionless signal amplitude of:
\begin{eqnarray}
2|f\tilde{h}(f)| & \sim & 5\times10^{-21}
	\left(\frac{M}{1\mathrm{kg}}\right)
	\left(\frac{v}{10\mathrm{m/s}}\right) \nonumber\\
& & \qquad\times\left(\frac{10\mathrm{Hz}}{f}\right)^3
	\left(\frac{5\mathrm{m}}{r_{\mathrm{min}}}\right)^3 \; .
\label{eq:colliding-weed-signal}
\end{eqnarray}
Thus even a ``typical'' 0.1kg weed at these speeds will produce a
signal-to-noise ratio of around 4 at 10~Hz in advanced LIGO
interferometers, which is in danger of being interpreted as a real
gravitational-wave event.  In a more extreme case, a 1kg tumbleweed
blown airborne by a strong 30m/s wind will produce a signal 100~times
higher than the noise at 10~Hz, which is easily detectable.

\begin{figure}[t]
\epsfig{file=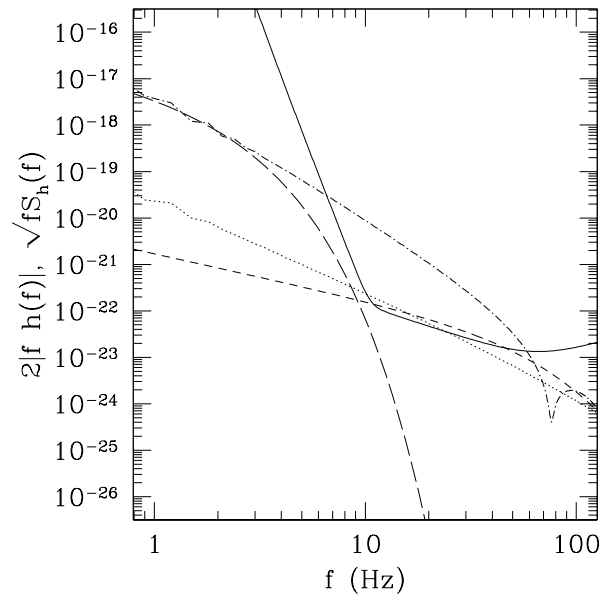,width=3.25in}
\caption{\label{fig:objects} Plot comparing dimensionless signal
strengths $2|f\tilde{h}(f)|$ of gravity gradients from airborne
objects with the dimensionless interferometer noise amplitude
$\sqrt{fS_h(f)}$, as a function of frequency.  The solid line is the
projected noise amplitude for advanced LIGO detectors.  The
short-dashed line is the signal from a .30~calibre bullet passing next
to the interferometer end station at 1000m/s, while the dotted line is
the signal from that same bullet colliding with the end station.  The
long-dashed line is the signal from a large 1kg tumbleweed passing
next to the end station at 30m/s, while the dot-dashed line is the
signal from that same tumbleweed colliding with the end station,
assuming that it compresses by 20cm on impact.  The tumbleweed mass
and speed are both near the upper end of the expected range,
representing the greatest danger of spurious signals in the detector;
evidently these tumbleweed signals \emph{can} be detected quite easily
by advanced LIGO detectors.}
\end{figure}

Fig.~\ref{fig:objects} shows the signal spectra for the objects
discussed above, plotted against the dimensionless noise amplitude
expected in advanced LIGO detectors.
Since tumbleweeds are a potential source of spurious detectable
events, one should consider ways to reduce the tumbleweed gravity
gradient noise.  Fortunately, the signal goes as
$r_{\mathrm{min}}^{-3}$, so a simple fence preventing the weeds from
approaching the end station should be sufficient.  A fence 30m out
from the end station will reduce the signal-to-noise ratio to 1 for
tumbleweed masses up to 1kg and speeds up to 30m/s, reducing the risk
of spurious events.

\section{Conclusions}
\label{s:gradient-conclusions}

This paper has studied two sources of background gravity gradient
noise, from infrasonic atmospheric pressure waves and from
wind-advected temperature perturbations, in order to determine whether
they constitute a limiting noise floor for interferometric
gravitational-wave detectors---in particular, for the ``advanced''
LIGO detectors projected in~\cite{Abramovici_A:1992}.  The paper also
analyzed two sources of gravity gradient signals, from transient
atmospheric shockwaves and from massive airborne bodies, to determine
whether they would constitute detectable spurious events in these
interferometers.  The following summarizes the results and suggests
possible further work that may need to be done.

Current estimates suggest that infrasonic pressure waves will not be a
significant source of gravity gradient noise, being over two orders of
magnitude below the advanced LIGO noise floor at 10~Hz.  Nonetheless,
these estimates are not based on actual noise measurements at an
interferometer site, so infrasound measurements at these sites are
recommended to confirm them.  Further empirical studies might also
analyze the specific effects of building shapes and of infrasound
coherence lengths on the noise spectrum, although these refinements
would likely only serve to reduce noise estimates above 15~Hz or so.

Wind-advected temperature perturbations, although the dominant source
of atmospheric density fluctuations, do not produce significant
high-frequency gravity gradient noise, due to the long times that any
particular pocket of warm or cool air spends in the vicinity of an
interferometer test mass.  A possible exception is when the airflow
forms vortices around the interferometer buildings, since this will
produce a noise spectrum peaked around the typical vortex circulation
frequencies near the test mass.  The current crude analysis of these
effects suggests that the noise is still an order of magnitude below
the advanced LIGO noise floor at 10~Hz even in the worst-case
scenarios, but the model could be improved significantly.  Numerical
models of the airflow and of temperature perturbations near an
interferometer building may be required to settle this issue
definitively.

Gravity gradients from atmospheric shockwaves are potentially serious
sources of spurious signals in interferometric gravitational-wave
detectors.  For instance, the sonic boom from a supersonic aircraft
overflying an advanced LIGO detector could produce signal-to-noise
ratios of several hundred.  Although such overflights are expected to
be rare or nonexistent, they point out the potential seriousness of
shocks from weaker or more distant sources, even if the signals are
several orders of magnitude smaller.  It is therefore
\emph{strongly} recommended that advanced interferometric detectors
include infrasonic detectors as environmental monitors.  Such sensors
could easily be employed to veto spurious atmospheric gravity-gradient
events.

Gravity gradients from wind-borne objects such as tumbleweeds are
another possible source of spurious events in gravitational-wave
detectors, if these objects are allowed to collide with the buildings
housing the interferometers.  Fences or other structures should be
used to keep these objects at least 30~metres from the test masses, in
order to eliminate the risk of spurious signals.

Obviously the number of things that can affect interferometric
detectors through gravity gradient forces is immense; I have
considered here only the few sources that I considered the most
worrisome.  I encourage other researchers to consider the implications
of this often-neglected effect.

\begin{acknowledgments}
This work was supported by NSF grant \mbox{PHY-9424337}.  I would like
to thank Kip Thorne for his help and support throughout this project,
as well as Brad Sturtevant, Brian Kern, and Scott Hughes for their
insightful discussions about atmospheric phenomena.  I am especially
grateful to Fred Raab for collecting the vital statistics of Hanford
tumbleweeds.
\end{acknowledgments}

\appendix
\section{The temperature noise spectrum}
\label{a:temperature-noise-spectrum}

This appendix presents a more rigorous mathematical derivation of
Eq.~(\ref{eq:h-noise}) used in
Sec.~\ref{s:atmospheric-temperature-perturbations}.

Consider a time-varying field of temperature perturbations $\delta
T(\mathbf{r},t)$ about some average temperature $T$.  This produces a
gravitational perturbation $g_x(t)=\int dV\, G\rho xr^{-3}(\delta
T/T)$, where $\rho$ is the average air density, and the $x$-axis is
along the interferometer arm.  The spectral density of gravity
gradient noise $S_g(|f|)$ is given by twice the Fourier transform of
the gravity autocorrelation $C_g(\tau)=\langle g(t)g(t+\tau)\rangle$.
Thus:
\begin{eqnarray}
S_g(|f|) & = & 2\left(\frac{G\rho}{T}\right)^2
		\int_{-\infty}^\infty d\tau
	\int dV \int dV' \frac{xx'}{r^3(r')^3} \nonumber\\
& & \langle\delta T(\mathbf{r},t)\delta T(\mathbf{r}',t+\tau)\rangle
	e^{2\pi if\tau} \; .
\label{eq:g-noise-integral}
\end{eqnarray}
The temperature noise measured at a point $\mathbf{r}_0$, on the other
hand, is given simply by:
\begin{equation}
\label{eq:t-noise-integral}
S_T(|f|) = \int_{-\infty}^\infty d\tau
	\langle\delta T(\mathbf{r}_0,t)\delta
	T(\mathbf{r}_0,t+\tau)\rangle
	e^{2\pi if\tau} \; .
\end{equation}

On sufficiently small scales, the temperature perturbations in the
Earth's turbulent boundary layer can be treated as homogeneous and
isotropic.  The expected squared temperature difference between two
points is then a function only of their separation: $\langle[\delta
T(\mathbf{r})-\delta T(\mathbf{r}+\Delta\mathbf{r})]^2\rangle
=D_T(||\Delta\mathbf{r}||)$.  The function $D_T(\Delta r)$ is called
the \emph{temperature structure function} of the atmosphere, and for
small $\Delta r$ reduces to a power law $D_T(r)=c_T^2\Delta r^p$.  If
a wind with speed $v$ blows these perturbations past a measuring
station, the temperature autocorrelation is $\langle\delta
T(\mathbf{r}_0,t)\delta T(\mathbf{r}_0,t+\tau)\rangle
=\sigma_T^2-(1/2)c_T^2(v\tau)^p$ for small $\tau$, where $\sigma_T^2$
is the mean squared temperature fluctuation.  This results in a
high-frequency power law tail:
\begin{equation}
\label{eq:t-noise}
S_T(|f|) = c_T^2 v^p (2\pi f)^{-(p+1)}\Gamma(p+1)\sin(p\pi/2) \; .
\end{equation}
Turbulent mixing theory, as well as micrometeorological measurements
of $S_T(|f|)$, show that the value of $p$ is normally 2/3,
characteristic of a type of turbulence known as Kolmogorov turbulence.
See, for example,~\cite{Busch_N:1972} for discussion of this type of
turbulent mixing, also~\cite{Coulman_C:1985} and references therein.

We are interested in $S_g(|f|)$, which is somewhat trickier to
calculate than $S_T(|f|)$, since it involves a correlation between
points separated in space as well as time.  However, chaotic
turbulence will almost certainly destroy high-frequency correlations
between widely separated points, so the high-frequency behaviour of
$S_g(|f|)$ will come from correlations between nearby points.  That
is, the high-frequency support of $\langle\delta T(\mathbf{r},t)\delta
T(\mathbf{r}',t+\tau)\rangle$ will come from those points
$\mathbf{r}'$ at time $t+\tau$ whose fluid elements were near
$\mathbf{r}$ at time $t$.

\begin{figure*}
\epsfig{file=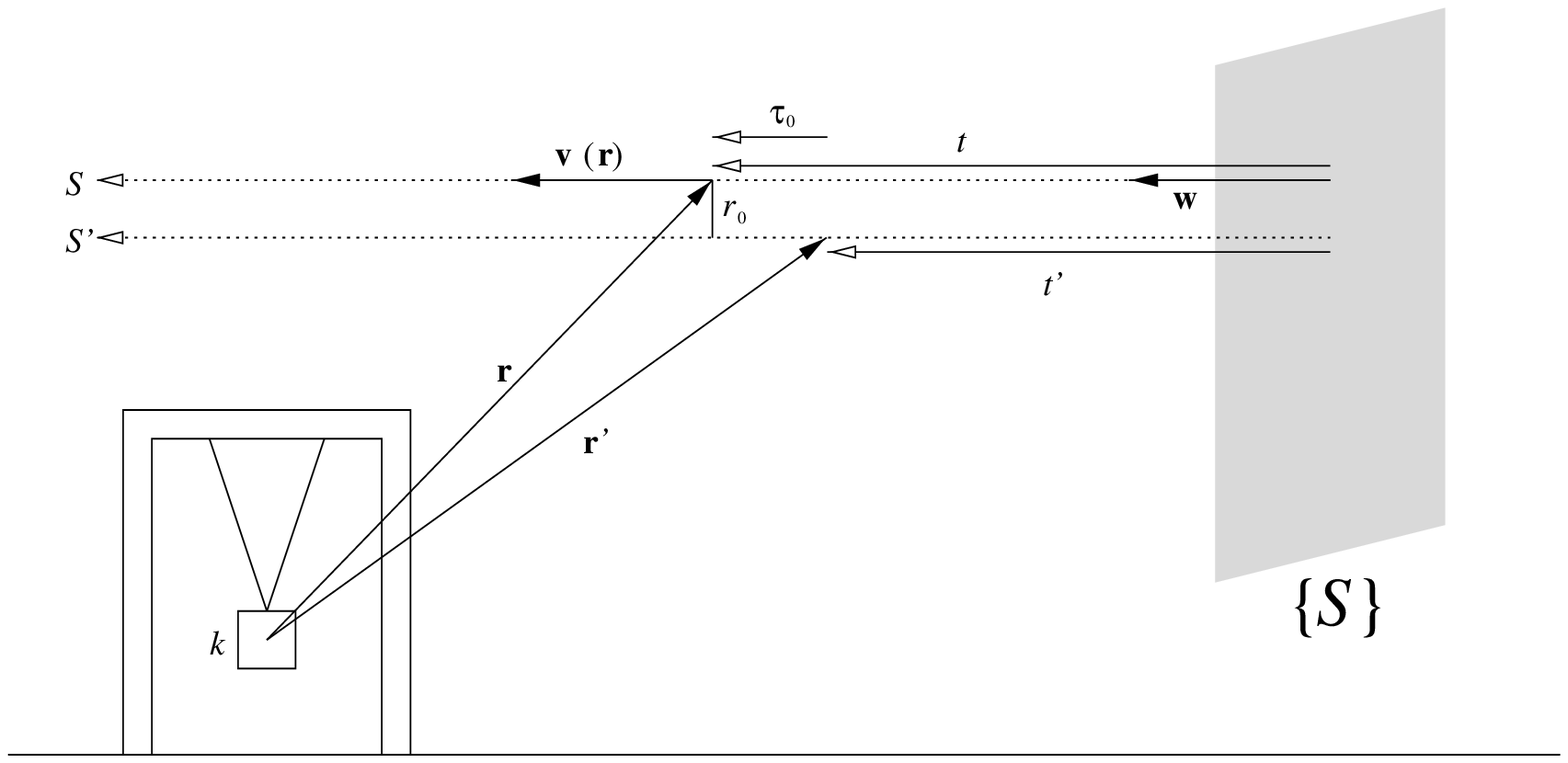,width=6.5in}
\caption{\label{fig:streamlines} Schematic showing the parameters used
to describe streamlines of air flowing past an interferometer test
mass.  $\{S\}$ denotes a plane intersecting all streamlines; $S$ and
$S'$ are two such streamlines passing through the points $\mathbf{r}$
and $\mathbf{r}'$, respectively, where the coordinate origin is
centered on the test mass $k$.  The distance between the streamlines,
measured at $\mathbf{r}$, is $r_0$.  The positions of the points
$\mathbf{r}$ and $\mathbf{r}'$ along the streamlines $S$ and $S'$ are
parameterized by the times $t$ and $t'$ that it would take for a
pocket of air to move from $\{S\}$ to $\mathbf{r}$ or $\mathbf{r}'$;
$\tau_0$ is the difference $t-t'$.  $\mathbf{v}(\mathbf{r})$ is the
wind velocity along $S$ through the point $\mathbf{r}$; $\mathbf{w}$
is the wind velocity through the plane $\{S\}$.}
\end{figure*}

Consider two fluid elements moving along paths $S$ and $S'$ passing
through $\mathbf{r}$ and $\mathbf{r}'$, respectively.  This is shown
schematically in Fig.~\ref{fig:streamlines}.  Let $r_0$ be the
distance from $\mathbf{r}$ to the nearest point on $S'$, and let
$\tau_0$ be the time it would take a pocket of air at $\mathbf{r}'$ to
be carried to this point on $S'$.  In order for these points
$\mathbf{r}$ and $\mathbf{r}'$ to contribute to the high-frequency
component of the spectrum, the distance $r_0$ must be fairly small, of
the order $v/f$ where $v$ is the wind speed past $\mathbf{r}$.  I
treat the streamlines as relatively straight over these scales, in
which case the temperature perturbations at $(\mathbf{r},t)$ and
$(\mathbf{r}',t+\tau)$ correspond to physical pockets of air separated
by a distance $\sqrt{r_0^2+v(\mathbf{r})^2(\tau-\tau_0)^2}$, for
$\tau$ near $\tau_0$.  Assuming that the $\tau$-dependence of the
correlation function is due entirely to this advection, the
correlation function can be written explicitly as:
\begin{equation}
\label{eq:t-correlation}
\langle\delta T(\mathbf{r},t)\delta T(\mathbf{r}',t+\tau)\rangle =
	\sigma_T^2 - \frac{c_T^2}{2}
	\left[r_0^2+v^2(\tau-\tau_0)^2\right]^{p/2} \; ,
\end{equation}
where typically $p\approx2/3$.

This term contains the entire $\tau$-dependence of the gravity
perturbation in Eq.~(\ref{eq:g-noise-integral}), so the first integral
I do is the Fourier integral over $\tau$.  This integral has the form
$\int_{-\infty}^\infty (\beta^2+x^2)^{\nu-1/2} e^{iax}\,dx
=2\pi^{-1/2}(2\beta/a)^\nu
\cos(\pi\nu)\Gamma(\nu+1/2)K_{-\nu}(a\beta)$.  Formally the integral
diverges for $\nu\geq1/2$, but the closed-form expression remains
approximately correct for large $a$ provided the integrand is cut off
smoothly for large $x\gg1/a$.  Physically this corresponds to the fact
that a smooth, large-scale cutoff of the temperature correlations in
Eq.~(\ref{eq:t-correlation}) will not affect the high-frequency
component of the temperature noise.  For horizontal winds near the
ground the spatial correlation function is cut off on horizontal
distance scales of $\sim50$ times the fluid elements' altitude $z$
above ground, giving a low-frequency cutoff around $\sim0.02v/z$
(Fig.~1.A4 of~\cite{Busch_N:1972}).  Typically this will be below the
relevant frequency range for interferometric detectors; I ignore it to
obtain pessimistic (upper-limit) noise estimates.  The high-frequency
noise tail is then:
\newpage
\begin{eqnarray}
S_g(|f|) & = & \left(\frac{G\rho}{T}\right)^2 \int dV \int dV'
	\bigg[\frac{xx'}{r^3(r')^3} \nonumber\\
& \times & c_T^2v^p(\sqrt{2}\pi f)^{-(p+1)}a(p)
	\left(\frac{2\pi fr_0}{v}\right)^{(p+1)/2} \nonumber\\
& & \times\; K_{(p+1)/2}(2\pi fr_0/v)e^{2\pi if\tau_0}\bigg] \; ,
\label{eq:g-noise-integral2}
\end{eqnarray}
where $a(p)=-2\pi^{-1/2}\cos(\pi[p+1]/2)\Gamma([p+2]/2)\approx0.873$
for $p=2/3$.

Next is the integral over $dV'$.  The exponential decay of the Bessel
function $K_{(p+1)/2}$ restricts the support of this integral to
values $r_0\alt v/2\pi f$, representing the fact that high-frequency
fluctuations can only arise from the rapid change of the spatial
correlation function over small lengthscales.  This range in $r_0$
defines a narrow bundle of streamlines $S'$ about the streamline $S$
passing through $\mathbf{r}$.  I assume that the size of this bundle
is less than the distance from the test mass to the bundle, so that
the values of $x'$ and $r'$ on a given $S'$ can be replaced with the
nearby values on $S$.  Now for points near $\mathbf{r}$ the volume
element $dV'$ can be written in terms of the new parameters $r_0$ and
$\tau_0$ as $dV'=2\pi r_0\,dr_0\,v(\mathbf{r})d\tau_0$.  Since the
airflow, being very subsonic, is nearly incompressible, the volume
element retains this form for all points along the bundle: if
$v(\mathbf{r}')$ decreases below $v(\mathbf{r})$, for instance, the
length element $v(\mathbf{r}')d\tau_0$ will decrease, but the
cross-section of the bundle (i.e., the relevant range of $2\pi
r_0dr_0$) will increase to compensate.  Plugging in this volume
element and integrating over $r_0$, one obtains:
\begin{eqnarray}
S_g(|f|) & = & \left(\frac{G\rho}{T}\right)^2 \int dV \int v\,d\tau_0
	\Bigg[\frac{xx'}{r^3(r')^3}c_T^2v^p(\pi f)^{-(p+1)}
\nonumber \\
& \times & a(p)2\pi\Gamma([p+3]/2)
	\left(\frac{v}{2\pi f}\right)^2 e^{2\pi if\tau_0}\Bigg] \; .
\label{eq:g-noise-integral3}
\end{eqnarray}

Let $t'$ be a new time coordinate denoting the time it takes for an
air pocket to reach $\mathbf{r}'$ from some fixed reference plane that
crosses all streamlines orthogonally, and $t$ be the corresponding
coordinate for the point $\mathbf{r}$.  Then $\tau_0$ is just $t'-t$,
and the volume element $dV$ can be written as $w\,dt\,dA$, where $dA$
is a cross-sectional area element on the reference plane, and $w$ is
the wind speed across that area element.  Plugging this in, and
ignoring the spatial separation between $S$ and $S'$, one obtains:
\begin{eqnarray}
S_g(|f|) & = & \left(\frac{G\rho}{T}\right)^2 c_T^2 (\pi f)^{-(p+3)}
	a(p)(\pi/2)\Gamma([p+3]/2) \nonumber\\
& \times & \int_{\{S\}} w\,dA
	\left(\int_{-\infty}^\infty \frac{x}{r^3}v^{p+3}
		e^{-2\pi ift}dt\right) \nonumber\\
& & \hspace{4em}\left(\int_{-\infty}^\infty \frac{x'}{(r')^3}
			e^{2\pi ift}dt'\right) \; ,
\label{eq:g-noise}
\end{eqnarray}
where I have used $\int_{\{S\}}$ to denote an integral over the entire
reference plane; i.e., over all streamlines $S$.

The noise in the gravitational-wave signal $h(t)$ due to the gradients
at a given test mass is $S_h(|f|)=(2\pi f)^{-4}S_g(|f|)/L^2$, where
$L$ is the interferometer arm length.  The noise at each test mass
adds incoherently to the total signal.  Combining these with
Eq.~(\ref{eq:g-noise}) yields the result given in
Eq.~(\ref{eq:h-noise}).


\begin{thebibliography}{10}
\expandafter\ifx\csname bibnamefont\endcsname\relax
  \def\bibnamefont#1{#1}\fi
\expandafter\ifx\csname bibfnamefont\endcsname\relax
  \def\bibfnamefont#1{#1}\fi
\expandafter\ifx\csname url\endcsname\relax
  \def\url#1{\texttt{#1}}\fi
\expandafter\ifx\csname urlprefix\endcsname\relax\def\urlprefix{URL }\fi
\providecommand{\bibinfo}[2]{#2}
\providecommand{\eprint}[2][]{\url{#2}}

\bibitem{Abramovici_A:1992}
\bibinfo{author}{\bibfnamefont{A.}~\bibnamefont{Abramovici}},
  \bibinfo{author}{\bibfnamefont{W.~E.} \bibnamefont{Althouse}},
  \bibinfo{author}{\bibfnamefont{R.~W.~P.} \bibnamefont{Drever}},
  \bibinfo{author}{\bibfnamefont{Y.}~\bibnamefont{G\"ursel}},
  \bibinfo{author}{\bibfnamefont{S.}~\bibnamefont{Kawamura}},
  \bibinfo{author}{\bibfnamefont{F.~J.} \bibnamefont{Raab}},
  \bibinfo{author}{\bibfnamefont{D.}~\bibnamefont{Shoemaker}},
  \bibinfo{author}{\bibfnamefont{L.}~\bibnamefont{Siewers}},
  \bibinfo{author}{\bibfnamefont{R.~E.} \bibnamefont{Spero}},
  \bibinfo{author}{\bibfnamefont{K.~S.} \bibnamefont{Thorne}},
  \bibinfo{author}{\bibfnamefont{R.~E.} \bibnamefont{Vogt}},
  \bibinfo{author}{\bibfnamefont{R.}~\bibnamefont{Weiss}}, \emph{et~al.},
  \bibinfo{journal}{Science}
  \textbf{\bibinfo{volume}{256}}(\bibinfo{number}{5055}), \bibinfo{pages}{325}
  (\bibinfo{year}{1992}).

\bibitem{Saulson_P:1984}
\bibinfo{author}{\bibfnamefont{P.~R.} \bibnamefont{Saulson}},
  \bibinfo{journal}{Phys. Rev. D.}
  \textbf{\bibinfo{volume}{30}}(\bibinfo{number}{4}), \bibinfo{pages}{732}
  (\bibinfo{year}{1984}).

\bibitem{Hughes_S:1998}
\bibinfo{author}{\bibfnamefont{S.~A.} \bibnamefont{Hughes}} \bibnamefont{and}
  \bibinfo{author}{\bibfnamefont{K.~S.} \bibnamefont{Thorne}},
  \bibinfo{journal}{Phys. Rev. D.}
  \textbf{\bibinfo{volume}{58}}(\bibinfo{number}{12}), \bibinfo{pages}{122002}
  (\bibinfo{year}{1998}), \eprint{gr-qc/9806018}.

\bibitem{Posmentier_E:1974}
\bibinfo{author}{\bibfnamefont{E.~S.} \bibnamefont{Posmentier}},
  \bibinfo{journal}{J. Geophys. Res.}
  \textbf{\bibinfo{volume}{79}}(\bibinfo{number}{12}), \bibinfo{pages}{1755}
  (\bibinfo{year}{1974}).

\bibitem{Busch_N:1972}
\bibinfo{author}{\bibfnamefont{N.~E.} \bibnamefont{Busch}}, in
  \emph{\bibinfo{booktitle}{Workshop on Micrometeorology}}, edited by
  \bibinfo{editor}{\bibfnamefont{D.~A.} \bibnamefont{Haugen}}
  (\bibinfo{publisher}{American Meteorological Society},
  \bibinfo{address}{Boston}, \bibinfo{year}{1972}), chap.~\bibinfo{chapter}{1},
  pp. \bibinfo{pages}{1--65}.

\bibitem{Coulman_C:1985}
\bibinfo{author}{\bibfnamefont{C.~E.} \bibnamefont{Coulman}},
  \bibinfo{journal}{Ann. Rev. Astron. Astrophys.}
  \textbf{\bibinfo{volume}{23}}, \bibinfo{pages}{19} (\bibinfo{year}{1985}).

\bibitem{Landau_L:1989}
\bibinfo{author}{\bibfnamefont{L.~D.} \bibnamefont{Landau}} \bibnamefont{and}
  \bibinfo{author}{\bibfnamefont{E.~M.} \bibnamefont{Lifshitz}},
  \emph{\bibinfo{title}{Fluid Mechanics}} (\bibinfo{publisher}{Pergamon Press},
  \bibinfo{address}{New York}, \bibinfo{year}{1959}).

\bibitem{Hayes_W:1971}
\bibinfo{author}{\bibfnamefont{W.~D.} \bibnamefont{Hayes}},
  \bibinfo{journal}{Ann. Rev. Fluid Mech.} \textbf{\bibinfo{volume}{3}},
  \bibinfo{pages}{269} (\bibinfo{year}{1971}).

\bibitem{Thorne_K:1999}
\bibinfo{author}{\bibfnamefont{K.~S.} \bibnamefont{Thorne}} \bibnamefont{and}
  \bibinfo{author}{\bibfnamefont{C.~J.} \bibnamefont{Winstein}},
  \bibinfo{journal}{Phys. Rev. D.}
  \textbf{\bibinfo{volume}{60}}(\bibinfo{number}{12}), \bibinfo{pages}{082001}
  (\bibinfo{year}{1999}), \eprint{gr-qc/9810016}.

\bibitem{Balachandran_N:1977}
\bibinfo{author}{\bibfnamefont{N.~K.} \bibnamefont{Balachandran}},
  \bibinfo{author}{\bibfnamefont{W.~L.} \bibnamefont{Donn}}, \bibnamefont{and}
  \bibinfo{author}{\bibfnamefont{D.~H.} \bibnamefont{Rind}},
  \bibinfo{journal}{Science} \textbf{\bibinfo{volume}{197}},
  \bibinfo{pages}{47} (\bibinfo{year}{1977}).

\bibitem{Witham_G:1974}
\bibinfo{author}{\bibfnamefont{G.~B.} \bibnamefont{Witham}},
  \emph{\bibinfo{title}{Linear and Nonlinear Waves}}
  (\bibinfo{publisher}{Wiley}, \bibinfo{address}{New York},
  \bibinfo{year}{1974}).

\bibitem{Raab_F:1999}
\bibinfo{author}{\bibfnamefont{F.}~\bibnamefont{Raab}},
  \bibinfo{howpublished}{private communication} (\bibinfo{year}{1999}).

\end{thebibliography}
\end{document}